\documentstyle[12pt]{article}
\begin{document}
\title{SCALE DEPENDENT DIMENSIONALITY}
\author{B.G. Sidharth$^*$\\
B.M. Birla Science Centre, Hyderabad 500 063 (India)}
\date{}
\maketitle
\footnotetext{$^*$E-mail:birlasc@hd1.vsnl.net.in}
\begin{abstract}
We argue that dimensionality is not absolute, but that it depends on the
scale of resolution, from the Planck to the macro scale.
\end{abstract}
\section{Introduction}
Is dimensionality dependent on the scale of resolution, or is it independent
of this scale. This question becomes relevant in the light of some recent
work (for example Cf.\cite{r1}). It has ofcourse been pointed out that the
spin half character of a collection of Fermions leads to the usual three
dimensionality of our space\cite{r2,r3}, while the spin half itself is
associated with the Compton wavelength as discussed in recent papers
(Cf. for example\cite{r4}). Further, it was argued that as we approach
the Compton scale, we encounter lower dimensionality\cite{r5,r6}. In this paper
we point out that indeed the dimensionality is scale dependent.
\section{Scale Dependence}
We first notice that at Planck scale $l_P$, we have
\begin{equation}
N^{3/4} l_P \sim R\label{e1}
\end{equation}
where $N \sim 10^{80}$ is the number of elementary particles and $R \sim 10^{28}cm$
the radius of the universe. This is not an empirical relation but rather can
be deduced on the basis of a fluctuational creation of particles scheme recently
discussed (Cf. for example\cite{r7,r8}). In this scheme, $\sqrt{N}$ particles
are fluctuationally created and this happens in the Compton time $\tau$ of a
typical elemental particle, a pion.\\
Further this corresponds to a fluctuational creation of $N^{1/4}$ Planck
particles as recently argued\cite{r9}, in the Planck time $\tau_P$. Indeed,
we have
\begin{equation}
\dot N \sim N^{1/4}/\tau_P \sim \sqrt{N}/\tau\label{e2}
\end{equation}
Equation (\ref{e2}) leads to (\ref{e1}).\\
(\ref{e1}) shows that at the Planck length, the fluctuational dimensionality
is $4/3$. Interestingly this is the dimension of a Koch curve and a coastline
\cite{r10}. With this dimensionality we should have
$$M \propto R^{4/3},$$
which indeed is true\cite{r11}.\\
At the Compton scale of resolution, we have\cite{r7}, as indeed can be deduced
from (\ref{e2}) the well known Eddington formula,
\begin{equation}
R \sim \sqrt{N} l\label{e3}
\end{equation}
(\ref{e3}) shows the two dimensional character at the Compton length. Indeed
as noted in the introduction three dimensionality is at scales much greater
than the Compton wavelength - as we approach the Compton wavelength we encounter
two dimensionality as can be seen from (\ref{e3}) - indeed this was the key
to explain puzzling characteristics of quarks including their fractional  charge
and handedness\cite{r6}.\\
Finally at scales $L \sim 10 cm$, we have
\begin{equation}
N^{1/3} L \sim R\label{e4}
\end{equation}
(\ref{e4}) shows up the usual three dimensionality.\\
Interestingly, if we take the typical elementary particle the pion, and consider
it successively as a $4/3$ dimensional object at the Planck scale, a two dimensional object
at the Compton scale and three dimensional at our macro scale, and consider
successive densities
$$\rho_P \sim m/(l_P)^{4/3}, \rho_\pi \sim m/l^2 \quad \mbox{and}\quad \rho \sim m/L^3,$$
we have,
$$M \sim \rho_P R^{4/3} \sim \rho_\pi R^2 \sim \rho R^3,$$
as required.

\end{document}